\documentclass[conference]{IEEEtran}
\IEEEoverridecommandlockouts
\usepackage{amsmath, amssymb, bm, cite, epsfig, psfrag, mathtools}
\usepackage[letterpaper,top=0.72in, bottom=1.05in, left=0.625in, right=0.625in]{geometry}
\usepackage{algorithm,algpseudocode}
\usepackage[utf8]{inputenc}
\usepackage[english]{babel}
\usepackage{xcolor}
\usepackage[figurename=Fig.]{caption}
\newtheorem{theorem}{Theorem}

\makeatletter
\renewcommand{\fnum@algorithm}{\fname@algorithm}
\makeatother

\def\BibTeX{{\rm B\kern-.05em{\sc i\kern-.025em b}\kern-.08em
		T\kern-.1667em\lower.7ex\hbox{E}\kern-.125emX}}

\begin{document}
	
	\title{Optimally Supporting IoT with Cell-Free Massive MIMO\\
	}
	
	\author{\IEEEauthorblockN{Hangsong Yan}
		\IEEEauthorblockA{\textit{NYU Wireless} \\
			New York, US\\
			hy942@nyu.edu}
		\and
		\IEEEauthorblockN{Alexei Ashikhmin}
		\IEEEauthorblockA{\textit{Bell Labs, Nokia} \\
			New Jersey, US \\
			alexei.ashikhmin@nokia-bell-labs.com}
		\and
		\IEEEauthorblockN{Hong Yang}
		\IEEEauthorblockA{\textit{Bell Labs, Nokia} \\
			New Jersey, US \\
			h.yang@nokia-bell-labs.com}}
	
	\maketitle
	
	\begin{abstract}
		We study internet of things (IoT) systems supported by cell-free (CF) massive MIMO (mMIMO) with optimal linear channel estimation. For the uplink, we consider optimal linear MIMO receiver and obtain an uplink SINR approximation involving only large-scale fading coefficients using random matrix (RM) theory. Using this approximation we design several max-min power control algorithms that incorporate power and rate weighting coefficients to achieve a target rate with high energy efficiency. For the downlink, we consider maximum ratio (MR) beamforming. Instead of solving a complex quasi-concave problem for downlink power control, we employ a neural network (NN) technique to obtain comparable power control with around 30 times reduction in computation time. For large networks we proposed a different NN based power control algorithm. This algorithm is sub-optimal, but its big advantage is that it is scalable.
	\end{abstract}
	
	
	\section{Introduction}\label{Introduction}
	Realization of internet of things (IoT) brings several new challenges, which include massive connectivity and very low device power budget requirements.
	In this paper, we propose to use Cell Free (CF) Massive multiple-input and multiple-output (mMIMO) to meet these challenges. A CF mMIMO system consists of many service antennas scattering throughout the entire intended coverage area. To allow active users identification, distinct, and therefore non-orthogonal, pilots are assigned to users, and linear minimum mean square error (LMMSE) is used for channel estimation. In the uplink (UL), we study the performance of optimal linear receiver, i.e., MMSE receiver. Other options like message passing based receivers \cite{Liu_Conver_Analysis} can also be considered. In the downlink (DL), we study the performance of maximum ratio (MR) beamforming. Our contributions can be summarized as follows. We derive an accurate, yet simple UL signal-to-noise-plus-interference ratio (SINR) approximation based on random matrix (RM) theory. We then obtain efficient and low complexity max-min power control algorithms based on this SINR approximation. Furthermore, we develop a target rate UL power control algorithm that vastly improves the device energy efficiency. In the DL, max-min power control for CF mMIMO has too large complexity. To reduce the complexity, we adopt a neuarl network (NN) approach. By predicting normalized transmit power for every access point (AP) under optimal max-min power
	control, DL max-min power control is converted from a high complexity quasi-concave problem to a low-complexity convex optimization problem. With the aid of NN prediction,
	we further develop a scalable power control algorithm that can work for very large areas with reasonably low complexity. Detail simulations are performed to validate our results.

	\section{System Model}\label{System Model}
	We consider an IoT system supported by CF mMIMO. We assume that $M$ APs and $\bar{K}$ things are uniformly distributed in a serving area.  At any given time, $K$ active things are randomly selected for service ($\bar{K} >> K$ and $M >> K$). We assume that active things are enumerated by $1,..., K$. 
	We also assume that orthogonal frequency-division multiplexing (OFDM) modulation is used and in what follows consider only one OFDM tone. In a given OFDM tone the channel coefficient $g_{mk}$ between $m$-th AP and $k$-th thing is
	\begin{equation}
	\label{eq:ch_coeffi}
	g_{mk} = \sqrt{\beta_{mk}}h_{mk}, m = 1,\ldots ,M, k = 1,\ldots ,K,
	\end{equation}
	where $\beta_{mk}$ are the large-scale fading coefficients, which include path loss and shadow fading, and $h_{mk}$ are small scale fading coefficients that are modeled as $i.i.d. \ \mathcal{CN}(0,1)$ random variables.
	All $\bar{K}$ users are assigned with distinct pilots,  which allow APs to identify active users. Since typically $\bar{K}$ is very large these pilots are unavoidably non-orthogonal. For example, the techniques from \cite{Liu_MassiveConnec_2018} and \cite{ Zhang_UAD_CE} may be used for this. 
	During the pilot transmission, pilots $\boldsymbol{\psi}_{k} \in \mathbb{C}^{\tau\times 1}, ||\boldsymbol{\psi}_{k}||_{2} = 1$, are synchronously transmitted by active things.  The signal received by the $m$-th AP is
	\begin{equation}
	\mathbf{y}_{m} = \sqrt{\tau\rho_{p}}\boldsymbol{\Psi}\mathbf{g}_{[m]} + \mathbf{w}_{m},
	\end{equation}
	where $\rho_{p}$ is the normalized signal-to-noise ratio (SNR) of each pilot symbol, $\mathbf{g}_{[m]} = [g_{m1},g_{m2},...,g_{mK}]^{T} \in \mathbb{C}^{K\times 1}$, $\boldsymbol{\Psi} = [\boldsymbol{\psi}_{1}\,\boldsymbol{\psi}_{2}\,...\,\boldsymbol{\psi}_{K}] \in \mathbb{C}^{\tau \times K}$,
	and $\mathbf{w}_m\in \mathbb{C}^{\tau \times 1}$ is the noise vector at $m$-th AP with $i.i.d.\ \mathcal{CN}(0,1)$ entries.
	
	APs use LMMSE channel estimation of $\mathbf{\bar{g}}_{m}$ and get
	\begin{equation}
	\begin{aligned}
	\label{eq:est_g}
	\hat{\mathbf{g}}_{[m]} =  \sqrt{\tau\rho_{p}}\mathbf{B}_{m}\boldsymbol{\Psi}^{H}(\tau\rho_{p}\boldsymbol{\Psi}\mathbf{B}_{m}\boldsymbol{\Psi}^{H} + \mathbf{I}_{\tau})^{-1}\mathbf{y}_{m},\\
	\hat{g}_{mk} = \sqrt{\tau\rho_{p}}\beta_{m,k}\boldsymbol{\psi}_{k}^{H}(\tau\rho_{p}\boldsymbol{\Psi}\mathbf{B}_{m}\boldsymbol{\Psi}^{H} + \mathbf{I}_{\tau})^{-1}\mathbf{y}_{m},
	\end{aligned}
	\end{equation}
	where $\mathbf{B}_{m} = \text{diag}\{[\beta_{m1},\beta_{m2},...,\beta_{mK}]\}$. Then the variance of the estimated channel coefficient $\hat{g}_{mk}$ is equal to
	
	\begin{equation}
	\begin{aligned}
	\label{eq:gamma}
	\gamma_{mk} \triangleq \mathbb{E}\{|\hat{g}_{mk}|^{2}\} = &\sqrt{\tau\rho_{p}}\beta_{mk}\boldsymbol{\psi}_{k}^{H}\mathbf{a}_{mk},
	\end{aligned}
	\end{equation}
	where $\mathbf{a}_{mk} = \sqrt{\tau\rho_{p}}\beta_{mk}(\tau\rho_{p}\boldsymbol{\Psi}\mathbf{B}_{m}\boldsymbol{\Psi}^{H} + \mathbf{I}_{\tau})^{-1}\boldsymbol{\psi}_{k}$. We denote by  $\tilde{g}_{mk}$ the channel coefficient estimation error.
	Note that $\tilde{g}_{mk}$ is uncorrelated with $\hat{g}_{mk}$ and that
	\begin{equation}
	\label{eq:est_err_g}
	\tilde{g}_{mk}\sim {\cal CN}(0,  \beta_{mk} - \gamma_{mk}).
	\end{equation}
	
	\section{Uplink Transmission}
	We consider UL with  MMSE MIMO receiver.
	\subsection{Uplink Data Transmission}
	Denote by $0\le \eta_{k} \le 1$  the power control coefficient for $k$-th thing.
	The received UL signals at the APs form the vector
	\begin{equation}
	\label{eq:rece_sig_UL}
	\mathbf{y}_{u} = \sqrt{\rho_{u}}\sum_{k=1}^{K}\sqrt{\eta_{k}}\mathbf{g}_{k}s_{k} + \mathbf{w}_{u},
	\end{equation}
	where $s_{k}$ is the symbol transmitted by $k$-th thing, $\rho_{u}$ is the normalized uplink SNR, $\mathbf{g}_{k} = [g_{1k}, g_{2k},..., g_{Mk}]^{T} \in \mathbb{C}^{M\times 1}$, and $\mathbf{w}_{u} \in \mathbb{C}^{M\times 1}$ is the noise vector with i.i.d. $\mathcal{CN}(0,1)$ entries.
	
	The central processing unit (CPU) of the CF mMIMO system estimates $s_k$ using linear MIMO receiver
	\begin{equation}
	\label{eq:decode_symbol}
	\begin{aligned}
	\hat{s}_{k} = \mathbf{v}_{k}^{H}\mathbf{y}_{u} =  \mathbf{v}_{k}^{H}\Big(&\sqrt{\rho_{u}\eta_{k}}\hat{\mathbf{g}}_{k}s_{k} + \sqrt{\rho_{u}}\sum_{k'\neq k}\sqrt{\eta_{k'}}\hat{\mathbf{g}}_{k'}s_{k'} +\\ &\sqrt{\rho_{u}}\sum_{k'=1}^{K}\sqrt{\eta_{k'}}\tilde{\mathbf{g}}_{k'}s_{k'} + \mathbf{w}_{u}\Big).
	\end{aligned}
	\end{equation}
	
	Based on (\ref{eq:est_err_g}) and (\ref{eq:decode_symbol}), the UL SINR expression for $k$-th data symbol is given as:
	\begin{equation}
	\label{eq:SINR_k_UL}
	\text{SINR}_{k}^{u}(\boldsymbol{\eta}) = \frac{\rho_{u}\eta_{k}\mathbf{v}_{k}^{H}\hat{\mathbf{g}}_{k}\hat{\mathbf{g}}_{k}^{H}\mathbf{v}_{k}}{\mathbf{v}_{k}^{H}(\rho_{u}\sum_{k'\neq k}\eta_{k'}\hat{\mathbf{g}}_{k'}\hat{\mathbf{g}}_{k'}^{H} + \mathbf{D})\mathbf{v}_{k}},
	\end{equation}
	where
	\begin{equation}
	\label{eq:def_D}
	\mathbf{D} = \rho_{u}\sum_{k'=1}^{K}\eta_{k'}(\mathbf{B}_{k'}-\boldsymbol{\Gamma}_{k'}) + \mathbf{I}_{M},
	\end{equation}
	$\mathbf{B}_{k'} \triangleq \text{diag}\{[\beta_{1k'}, \beta_{2k'},..., \beta_{Mk'}]\}$, and $\mathbf{\Gamma}_{k'} \triangleq \text{diag}\{[\gamma_{1k'}, \gamma_{2k'},...,\gamma_{Mk'}]\}$. Based on (\ref{eq:decode_symbol}) and (\ref{eq:SINR_k_UL}), and using Rayleigh-Ritz theorem, we obtain that the optimal choise of $\mathbf{v}_{k}$, i.e., LMMSE MIMO receiver, is
	\begin{equation}
	\label{eq:decoder_MMSE}
	\mathbf{v}_{k}^{\text{MMSE}} = \sqrt{\rho_{u}\eta_{k}}(\rho_{u}\sum_{k'=1}^{K}\eta_{k'}\hat{\mathbf{g}}_{k'}\hat{\mathbf{g}}_{k'}^{H} + \mathbf{D})^{-1}\hat{\mathbf{g}}_{k}.
	\end{equation}
	Substituting (\ref{eq:decoder_MMSE}) into (9), we then obtain the corresponding SINR expression:
	\begin{equation}
	\begin{aligned}
	\label{eq:SINR_k_UL_MMSE}
	&\text{SINR}_{k}^{u, \text{MMSE}}(\boldsymbol{\eta}) = \\ &\frac{\rho_{u}\eta_{k}\hat{\mathbf{g}}_{k}^{H}\left(\rho_{u}\sum_{k'=1}^{K}\eta_{k'}\hat{\mathbf{g}}_{k'}\hat{\mathbf{g}}_{k'}^{H} + \mathbf{D}\right)^{-1}\hat{\mathbf{g}}_{k}}{1-\rho_{u}\eta_{k}\hat{\mathbf{g}}_{k}^{H}\left(\rho_{u}\sum_{k'=1}^{K}\eta_{k'}\hat{\mathbf{g}}_{k'}\hat{\mathbf{g}}_{k'}^{H} + \mathbf{D}\right)^{-1}\hat{\mathbf{g}}_{k}}.
	\end{aligned}
	\end{equation}
	
	\subsection{Random Matrix Approximation of SINR}
	Using results of RM theory from \cite{Hoydis_13_UL/DL}, \cite{Wagner_12_MISO} we obtain the following approximation of SINR:
	\begin{equation}
	\label{eq:RMT_Approx1}
	\text{SINR}_{k}^{u,\text{AP}}(\boldsymbol{\eta}) = \frac{\rho_{u}\eta_{k}\text{tr}\boldsymbol{\Gamma}_{k}\mathbf{T}}{M},
	\end{equation}
	where
	\begin{equation}
	\label{eq:T_Approx1}
	\mathbf{T} = (\frac{\rho_{u}}{M}\sum_{k'=1}^{K}\frac{\eta_{k'}\mathbf{\Gamma}_{k'}}{1 + e_{k'}} + \frac{\mathbf{D}}{M})^{-1},
	\end{equation}
	\begin{equation}
	\label{eq:e_k}
	e_{k'} = \lim_{t \rightarrow \infty} e_{k'}^{(t)}\ \text{with}\ e_{k'}^{(0)} = M, \;\;\forall k',
	\end{equation}
	\begin{equation}
	\label{eq:e_k_t}
	e_{k'}^{(t)} = \frac{\rho_{u}\eta_{k'}}{M}\text{tr}\mathbf{\Gamma}_{k'}(\frac{\rho_{u}}{M}\sum_{j=1}^{K}\frac{\eta_{j}\mathbf{\Gamma}_{j}}{1 + e_{j}^{(t-1)}} + \frac{\mathbf{D}}{M})^{-1}, \forall k'.
	\end{equation}

	The sketch of our derivations is the following. We first define $\boldsymbol{\Lambda} = \rho_{u}\sum_{k'=1}^{K}\eta_{k'}\hat{\mathbf{g}}_{k'}\hat{\mathbf{g}}_{k'}^{H} + \mathbf{D}$, $\boldsymbol{\Lambda}_{k} = \rho_{u}\sum_{k'\neq k}^{K}\eta_{k'}\hat{\mathbf{g}}_{k'}\hat{\mathbf{g}}_{k'}^{H} + \mathbf{D}$. Using matrix inversion lemma \cite[lemma 2.2]{silverstein1995empirical}), the term $\rho_{u}\eta_{k}\hat{\mathbf{g}}_{k}^{H}\boldsymbol{\Lambda}^{-1}\hat{\mathbf{g}}_{k}$ in (\ref{eq:SINR_k_UL_MMSE}) is given as  
	\begin{equation}
	\label{eq:lemma1}
	\rho_{u}\eta_{k}\hat{\mathbf{g}}_{k}^{H}\boldsymbol{\Lambda}^{-1}\hat{\mathbf{g}}_{k} = \frac{\rho_{u}\eta_{k}\hat{\mathbf{g}}_{k}^{H}\boldsymbol{\Lambda}_{k}^{-1}\hat{\mathbf{g}}_{k}}{1 + \rho_{u}\eta_{k}\hat{\mathbf{g}}_{k}^{H}\boldsymbol{\Lambda}_{k}^{-1}\hat{\mathbf{g}}_{k}}.
	\end{equation}
	Then using \cite[lemma 4]{Wagner_12_MISO} for $\rho_{u}\eta_{k}\hat{\mathbf{g}}_{k}^{H}\boldsymbol{\Lambda}_{k}^{-1}\hat{\mathbf{g}}_{k}$, we obtain:
	\begin{equation}
	\label{eq:lemma2}
	\rho_{u}\eta_{k}\hat{\mathbf{g}}_{k}^{H}\boldsymbol{\Lambda}_{k}^{-1}\hat{\mathbf{g}}_{k} - \rho_{u}\eta_{k}\text{tr}\mathbf{\Gamma}_{k}\boldsymbol{\Lambda}_{k}^{-1}\xrightarrow[M \rightarrow \infty]{\text{a.s.}} 0.
	\end{equation}
	By further applying \cite[lemma 6]{Wagner_12_MISO} to $\rho_{u}\eta_{k}\text{tr}\mathbf{\Gamma}_{k}\mathbf{\Lambda}_{k}^{-1}$, we obtain:
	\begin{equation}
	\rho_{u}\eta_{k}\text{tr}\mathbf{\Gamma}_{k}\mathbf{\Lambda}_{k}^{-1} - \rho_{u}\eta_{k}\text{tr}\mathbf{\Gamma}_{k}\mathbf{\Lambda}^{-1} \xrightarrow[M \rightarrow \infty]{\text{a.s.}} 0.
	\end{equation}
	Applying \cite[Theorem 1]{Hoydis_13_UL/DL} to $\rho_{u}\eta_{k}\text{tr}\mathbf{\Gamma}_{k}\mathbf{\Lambda}^{-1}$, we then obtain
	\begin{equation}
	\rho_{u}\eta_{k}\text{tr}\mathbf{\Gamma}_{k}\mathbf{\Lambda}^{-1} - \frac{\rho_{u}\eta_{k}}{M}\text{tr}\mathbf{\Gamma}_{k}\mathbf{T} \xrightarrow[M \rightarrow \infty]{\text{a.s.}} 0.
	\end{equation}
	where $\mathbf{T}$ is defined in (\ref{eq:T_Approx1}). Then $\rho_{u}\eta_{k}\hat{\mathbf{g}}_{k}^{H}\boldsymbol{\Lambda}_{k}^{-1}\hat{\mathbf{g}}_{k}$ is substituted by $\frac{\rho_{u}\eta_{k}}{M}\text{tr}\mathbf{\Gamma}_{k}\mathbf{T}$ in (\ref{eq:lemma1}), and (\ref{eq:lemma1}) is further substituted into (\ref{eq:SINR_k_UL_MMSE}) to obtain the RM Approximation in (\ref{eq:RMT_Approx1}). 
	Note that \cite[Theorem 1]{Hoydis_13_UL/DL} requires that $\hat{\mathbf{g}}_{k'},\, \forall k'$ be independent. The following Theorem shows that asymptotically this is the case. 
	\begin{theorem}\cite{Yan2020scalable}
		\label{Theorem 1}
		For $\hat{g}_{mk}$ and $\hat{g}_{ml}$ defined in (\ref{eq:est_g}) we have 
		
		\begin{equation}
		\label{eq:theorem_1}
		\text{Cov}[\hat{g}_{mk},\hat{g}_{ml}] \xrightarrow[\tau \rightarrow \infty]{\text{a.s.}} 0 \ \text{for}\ k,l = 1,...,K\ \text{and}\ k\neq l.
		\end{equation}
	\end{theorem}
	
	\section{ Uplink Power Control}
	We first consider an iterative max-min power control algorithm based on exact SINR given in (\ref{eq:SINR_k_UL_MMSE}).
	In the algorithm, a rate weighting vector $\mathbf{u} = [u_{1}, u_{2},..., u_{K}]^{T} \in \mathbb{R}_{0^{+}}^{K}$
	is used.
	Vector $\mathbf{u}$ can be used to drop some things under poor serving condition by assigning very small weights to these things. We also use a power weighting vector $\boldsymbol{\nu} = [\nu_{1}, \nu_{2}, ..., \nu_{K}]^{T} \in \mathbb{R}_{0^{+}}^{K}$. So,  the weighted normalized maximum transmit power of $k$-th thing is defined as $\rho'_{u,k} \triangleq \rho_{u}\nu_{k}$. We define matrix $\mathbf{J}_k, k = 1, 2,..., K,$ as
	\begin{equation}
	\label{eq:J_matrix}
	\mathbf{J}_{k} = \hat{\mathbf{g}}_{k}\hat{\mathbf{g}}_{k}^{H} + \mathbf{B}_{k} - \boldsymbol{\Gamma}_{k}.
	\end{equation}
	The algorithm is given in Algorithm \ref{Algorithm 1}.  The convergence proof is omitted due to page limit.
	\begin{algorithm}
		\caption{{\bf 1} Max-min Power Control - Exact SINR}
		\label{Algorithm 1}
		\begin{algorithmic}[1]
			\item Initialize vector $\mathbf{u}$ and $\boldsymbol{\nu}$ with predefined setting. Initialize $\eta_{k}^{(0)} = 1, \forall k$, and $d_{k}^{(0)}, \forall k$ as
			\begin{equation}
			\label{eq:d_k_0}
			d_{k}^{(0)} = \rho'_{u,k}\hat{\mathbf{g}}_{k}^{H}(\sum_{k'=1}^{K}\rho'_{u,k'}\eta_{k'}^{(0)}\mathbf{J}_{k'} + \mathbf{I}_{M})^{-1}\hat{\mathbf{g}}_{k}.
			\end{equation}
			Set $n = 0$ and choose a tolerance $\epsilon > 0$.
			\State Compute $\alpha = \min_{k'}(d_{k'}^{(n)}/u_{k'}), k' = 1,..., K$ and update $d_{k}^{(n+1)}, \forall k$ as below
			\begin{equation}
			\begin{aligned}
			\label{eq:d_updation}
			d_{k}^{(n+1)} = \rho'_{u,k}\hat{\mathbf{g}}_{k}^{H}(\alpha\sum_{k'=1}^{K}\frac{\rho'_{u,k'}u_{k'}}{d_{k'}^{(n)}}\mathbf{J}_{k'} + \mathbf{I}_{M})^{-1}\hat{\mathbf{g}}_{k}.
			\end{aligned}
			\end{equation}
			\State Stop if $\max_{k}|d_{k}^{(n+1)} - d_{k}^{(n)}|\leq \epsilon, k = 1,..., K$, set $d_{k} = d_{k}^{(n+1)}, \forall k$, and the power control coefficients, $\eta_{k}, \forall k$ are given by
			\begin{equation}
			\label{eq:eta_1}
			\eta_{k} = \frac{\min_{k'}(d_{k'}/u_{k'})}{d_{k}/u_{k}}, \,  k' = 1,...,K.
			\end{equation}
			Otherwise, set $n = n + 1$ and go to step 2.
		\end{algorithmic}
	\end{algorithm}
	
	Remark: if all the things are required to achieve the same rate, $u_{k} = 1/\sqrt{K}, \forall k$.
	
	The disadvantage of the above algorithm is its high complexity (see below), and that it involves small-scale channel coefficients $h_{mk}$ and therefore requires frequent power updates, which can be very undesirable for low cost and low power sensors. For this reason we
	propose Algorithm \ref{Algorithm 2}  based on our RM Approximation. 
	This algorithm involves only large-scale fading coefficients, and therefore power control coefficients can be updated in a much slower rate. The convergence proof is omitted due to page limit.
	
	In IoT systems things will use energy harvesting and/or infrequently replaced batteries. Thus, high energy efficiency is very important. We define the energy efficiency of an UL IoT system by
	\begin{equation}
	\label{eq:EE}
	E_{u} = \frac{\sum_{k = 1}^{K}R_{k}^{u}}{P_{u}\sum_{k=1}^{K}\eta_{k}},
	\end{equation}
	where $R_{k}^{u}$ is the UL achievable rate for $k$-th thing, $P_{u}$ is the maximum transmit power for each thing in data transmission.
	
	For achieving high energy efficiency we propose power control algorithms in which
	each thing achieves a target rate with high energy efficiency. We propose Algorithm 3 based on exact SINR (\ref{eq:SINR_k_UL_MMSE}) and Algorithm 4 based on RM approximation. Algorithm 4 is presented below. Note that $u_g$ and $u_p$ in step 4 of Algorithm 4 are rate weighting coefficients assigned to good and poor things, respectively. Poor things are things whose data rates under full power transmission are less than the predefined target rate. Details of Algorithm 3 and the convergence proofs for Algorithm 3 and 4 are omitted due to page limits, but can be found in \cite{Yan2020scalable}. 
	
	To compare the complexity of these algorithms we first note that in Algorithm 1 the computation of (\ref{eq:SINR_k_UL_MMSE}) involves inversion of an $M \times M$ non-sparse matrix whose complexity
	is $\sim\mathcal{O}(M^{3})$. On the other hand, matrix $\mathbf{T}$ is a diagonal matrix, so the computation complexity of (\ref{eq:T_update}) in Algorithms 2 and 4 is $\sim\mathcal{O}(MK)$.

	\begin{algorithm}
		\caption{{\bf 2} Max-min Power Control - RM SINR}
		\label{Algorithm 2}
		\begin{algorithmic}[1]
			\item Initialize $\mathbf{u}$, $\boldsymbol{\nu}$ as predetermined, $\eta_{k}^{(0)} = 1, \forall k$, $\mathbf{D}^{(0)} = \sum_{k'=1}^{K}\rho'_{u,k'}\eta_{k'}^{(0)}(\mathbf{B}_{k'}-\boldsymbol{\Gamma}_{k'}) + \mathbf{I}_{M}$. Initialize $\mathbf{T}^{(0)} = (\frac{1}{M}\sum_{k'=1}^{K}\frac{\eta_{k'}^{(0)}\rho'_{u,k'}\mathbf{\Gamma}_{k'}}{1+e_{k'}} + \frac{\mathbf{D}^{(0)}}{M})^{-1}$ where $e_{k'}, \forall k'$ are computed by (\ref{eq:e_k}) and (\ref{eq:e_k_v_t}) below
			\begin{equation}
			\small
			\label{eq:e_k_v_t}
			e_{k'}^{(t)} =  \frac{\rho'_{u,k'}\eta_{k'}^{(0)}}{M}\text{tr}\mathbf{\Gamma}_{k'}(\frac{1}{M}\sum_{j=1}^{K}\frac{\eta_{j}^{(0)}\rho'_{u,j}\mathbf{\Gamma}_{j}}{1 + e_{j}^{(t-1)}} + \frac{\mathbf{D}^{(0)}}{M})^{-1}.
			\end{equation}
			Set $n = 0$ and choose a tolerance $\epsilon > 0$.
			
			\item Compute $\alpha = \min_{k'}\text{tr}(\nu_{k'}\mathbf{\Gamma}_{k'}\mathbf{T}^{(n)})/u_{k'}, k' = 1,..., K$, Update $\mathbf{T}^{(n+1)}$ as
			\begin{equation}
			\small
			\begin{aligned}
			\label{eq:T_update}
			&\mathbf{T}^{(n+1)} = (\frac{\alpha}{M}\sum_{k'=1}^{K}\frac{\rho_{u}u_{k'}}{\text{tr}\mathbf{\Gamma}_{k'}\mathbf{T}^{(n)}}(\mathbf{B}_{k'} - \frac{\xi_{k'}\boldsymbol{\Gamma}_{k'}}{1 + \xi_{k'}}) + \frac{\mathbf{I}_{M}}{M})^{-1},
			\end{aligned}
			\end{equation}
			where $\xi_{k'} = \frac{\rho_u \alpha u_{k'}}{M}$.
			
			\item Stop if $||\mathbf{T}^{(n+1)} - \mathbf{T}^{(n)}|| \leq \epsilon$. Set $\mathbf{T} = \mathbf{T}^{(n+1)}$ and the power control coefficients, $\eta_{k}, \forall k$ are given by
			\begin{equation}
			\label{eq:eta_al_AP1}
			\eta_{k} = \frac{\min_{k'}\text{tr}(\nu_{k'}\mathbf{\Gamma}_{k'}\mathbf{T})/u_{k'}}{\text{tr}(\nu_{k}\mathbf{\Gamma}_{k}\mathbf{T})/u_{k}}, k' = 1,..., K.
			\end{equation}
			Otherwise, set $n = n + 1$ and go to step 2.
		\end{algorithmic}	
	\end{algorithm}

	\begin{algorithm}
		\caption{{\bf 4} Target Rate Power Control - RM SINR}
		\label{Algorithm 4}
		\begin{algorithmic}[1]
			\item Initialize $\boldsymbol{\nu}$, $\eta_{k}^{(0)}, \forall k$, $\mathbf{D}^{(0)}$ and $\mathbf{T}^{(0)}$ as in step 1 of Algorithm \ref{Algorithm 2}. With a target SINR denoted as $S_{t}$, compute $\alpha = S_{t}M/\rho_{u}$. Set $n = 0$ and choose a tolerance $\epsilon > 0$.
			
			\item Update $\mathbf{T}^{(n+1)}$ using (\ref{eq:T_update}) where $\rho_{u}u_{k'}, \forall k'$ are substituted by $\rho_{u}$ and $\xi_{k'} = S_t, \forall k'$.
			
			\item Stop if $||\mathbf{T}^{(n+1)} - \mathbf{T}^{(n)}|| < \epsilon$ and set $\mathbf{T} = \mathbf{T}^{(n+1)}$. Otherwise, set $n = n + 1$ and go to step 2.
			
			\item Compute the power coefficients $\eta_{k} = \alpha/(\text{tr}\nu_{k}\mathbf{\Gamma}_{k}\mathbf{T}), \forall k$. If  $0 \leq \eta_{k} \leq 1, \forall k$ Algorithm 4 ends. Otherwise, initialize $\eta_{k}^{(0)}, \forall k$, $\mathbf{D}^{(0)}$ and $\mathbf{T}^{(0)}$ as in step 1 of Algorithm \ref{Algorithm 2}. Assign the value of each element of vector $\mathbf{u}$ by $u_{g}$ or $u_{p}$ according to the per-thing rate under full power case. Set $n = 0, \alpha=\alpha/u_{g}$, and go to step 5.
			
			\item Update $\mathbf{T}^{(n+1)}$ using (\ref{eq:T_update}) where $\xi_{k'} = \frac{S_t u_{k'}}{u_g}$. 
			
			\item Stop if $||\mathbf{T}^{(n+1)} - \mathbf{T}^{n}|| \leq \epsilon$, set $\mathbf{T} = \mathbf{T}^{(n+1)}$, and compute  $\eta_{k} = \alpha u_{k}/(\text{tr}\nu_{k}\mathbf{\Gamma}_{k}\mathbf{T}), \forall k$. Otherwise, set $n = n + 1$ and go to step 5.	
		\end{algorithmic}
	\end{algorithm}

	\section{Uplink Simulation Results}\label{Uplink Simulation Results}
	
	We consider networks where $M$ APs and $K$ things are uniformly distributed in a $D \times D \ m^{2}$ square serving area, which is wrapped around to avoid boundary effects. We use 
	
	\begin{equation}
	\label{eq:large_fading}
	\beta_{mk} = \text{PL}_{mk}\text{SF}_{mk}, \text{with}\ \text{SF}_{mk} = 10^{\frac{\sigma_{\text{sh}}z_{mk}}{10}},
	\end{equation}
	where $\text{PL}_{mk}$ is  the path loss and $\text{SF}_{mk}$ is the shadow fading with $z_{mk} \sim \mathcal{N}(0,1)$. The path loss is generated as in \cite{Ngo_17_Cellfree} where a three-slope model \cite{Tang_01_Pathloss} and the Hata-Cost 231 propagation model \cite{3GPP_ETSI} are used. Shadow fading coefficients are generated as in \cite{Wang08_Joint_Shadow}. We use $B=20$ MHz  bandwidth and $f=1.9$ GHz carrier frequency. As a performance measure, we use achievable rate defined by 
	\begin{equation}
	\label{eq:exact_throughput}
	U_{k}^{u,\text{MMSE}} = B((\tau_c - \tau)/(2\tau_c))R_{k}^{u, \text{MMSE}},
	\vspace{-0.3mm}
	\end{equation}
	\begin{equation}
	\label{eq:exact_rate}
	R_{k}^{u,\text{MMSE}} = \mathbb{E}\left[\log_{2}\left(1 + \text{SINR}_{k}^{u,\text{MMSE}}\right)\right],
	\end{equation}
	where $\tau_c$ is the length of coherence interval measured in OFDM symbols, the expectation in (\ref{eq:exact_rate}) is over small-scale fading. The  rate of $k$-th thing obtained via the RM Approximation is given by $R_{k}^{u,\text{AP}} = \log_{2}(1 + \text{SINR}_{k}^{u,\text{AP}})$.
	
\begin{figure}[th]
		\vspace{-5mm}
		\begin{center}
			\centering \scalebox{0.9}{\includegraphics[width=\columnwidth]{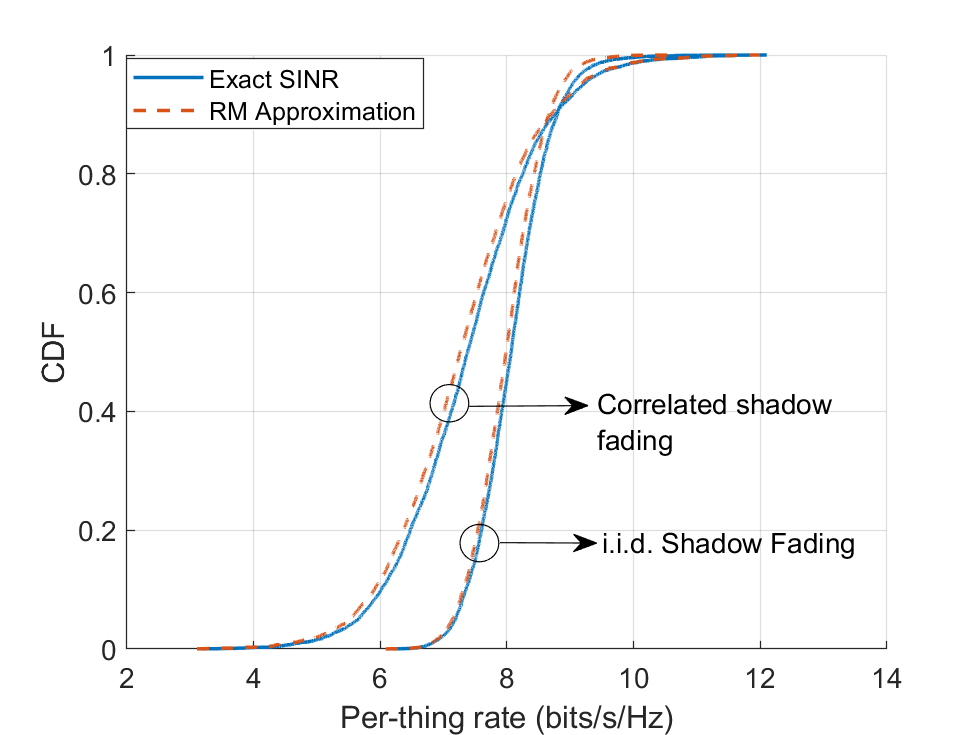}}\caption{$M = 1024$, $K = 256$, $\tau = 256$, $P_{u} = 20\ \text{mW}$, and area = 1 $\text{km}^{2}$
			}\label{fig:AP_Exact_Comparison}
		\end{center}
	\end{figure}

	
	Fig. \ref{fig:AP_Exact_Comparison} demonstrates that the RM Approximation of per-thing rate is quite accurate for both correlated and i.i.d. shadow fading. 
	
\begin{figure}
		\begin{center}
			\includegraphics
			[width=0.5\textwidth]{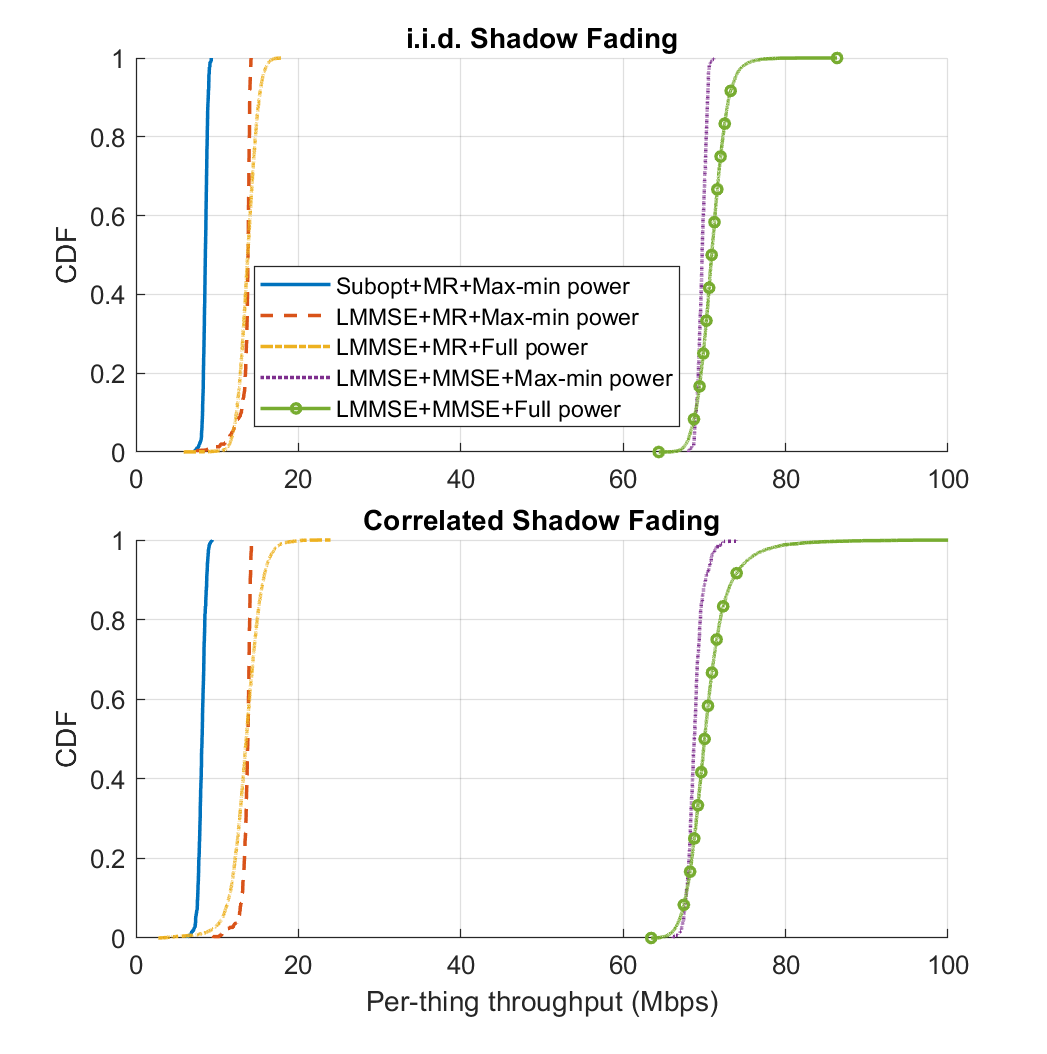} 
			\caption{$M = 128$, $K = 40$, $\tau = 60$, $P_{u} = 20\ \text{mW}$, and area = 0.01 $\text{km}^{2}$}
			\label{fig:Opt_subopt_max_min_comparison}
		\end{center}
	\end{figure}

	In Fig. \ref{fig:Opt_subopt_max_min_comparison}, `Subopt' denotes the sub-optimal channel estimation used in \cite{Ngo_17_Cellfree}. We see that LMMSE channel estimation gives a visible gain over the sub-optimal channel estimation, and that MMSE MIMO receiver gives  7-fold improvement over MR MIMO receiver for both i.i.d and correlated shadow fading cases. At the same time we do not observe any significant gain from the transmit power optimization. 
	\begin{figure}
		\begin{center}
			\centering \scalebox{1}{\includegraphics[width=\columnwidth]{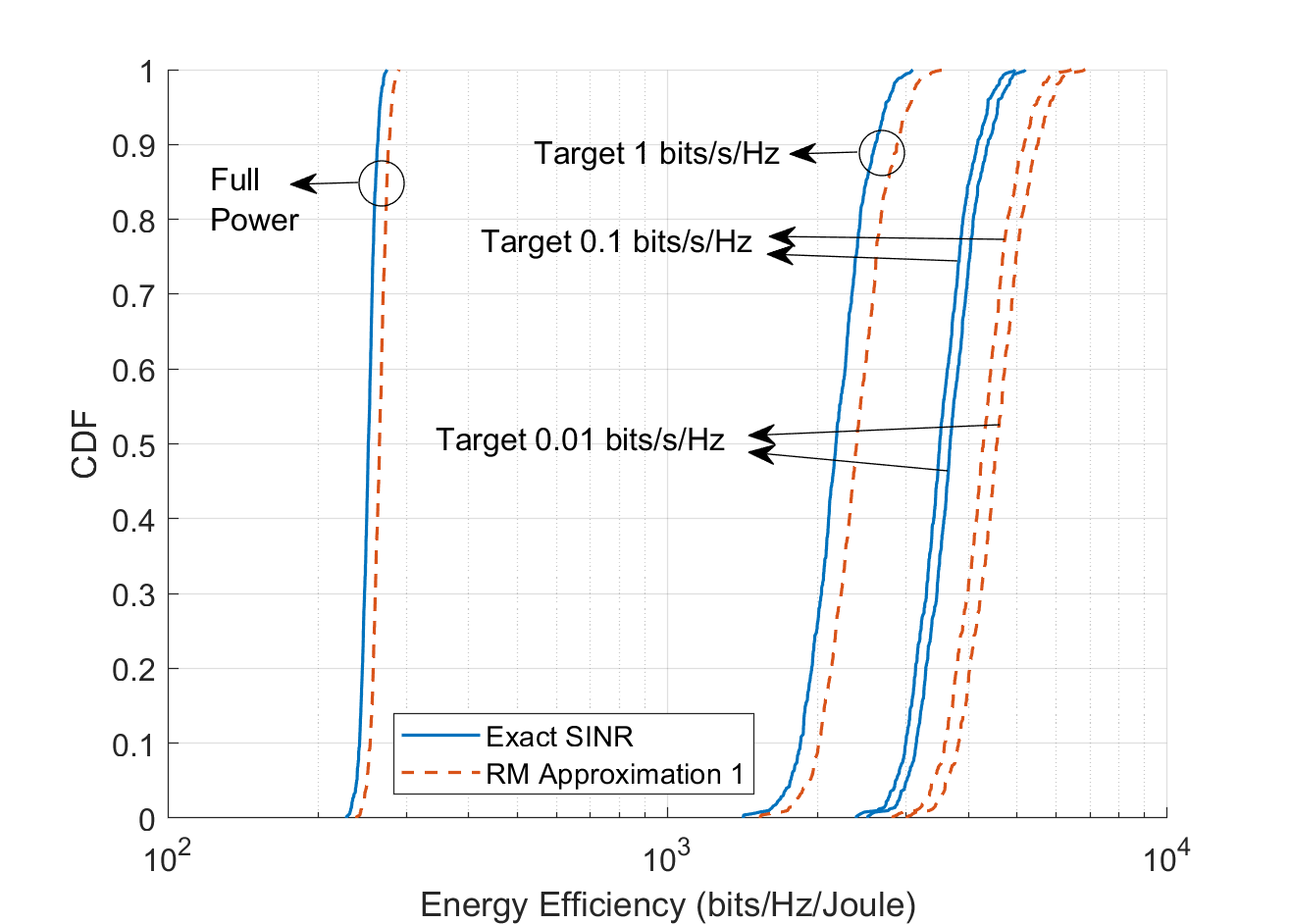}}
			\caption{$M = 160$, $K = 40$, $\tau = 40$, $P_{u} = 20\ \text{mW}$, and area = 1 $\text{km}^{2}$}
			\label{fig:EE_Comparison}
		\end{center}
	\end{figure}
	
	However, Fig. \ref{fig:EE_Comparison} demonstrates that transmit power optimization gives a very significant, up to 17 times, energy efficiency improvement. It also shows that the RM approximation is reasonably accurate in the case of optimized power coefficients.

	\section{Downlink Transmission}
	We consider DL transmission with LMMSE  channel estimation and MR MIMO precoding. 
	The received signal by the $k$-th thing is
	\begin{equation}\label{eq:y_d_k}
	y_{k}^{d} = \sqrt{\rho_{d}}\sum_{m=1}^{M}\sqrt{\eta_{mk}} \hat{g}_{mk}^{*}g_{mk} s_{k} + w_k,
	\end{equation}
	where  $\eta_{mk},\ m = 1,...,M,\ k = 1,..., K$ are DL power control coefficients and $w_k \sim \mathcal{CN}(0, 1)$ is the noise component at $k$-th thing. 
	The closed-form SINR expression (\ref{eq:SINR_d_k_radom}) was derived in \cite{Rao_Internet_2019} and \cite{Rao_cellfree_2018}.
	
	\begin{figure*}
		\begin{equation}
		\label{eq:SINR_d_k_radom}
		\begin{aligned}
		{\small 
			\text{SINR}_{k}^{\text{IoT}} = \frac{\rho_{d}(\sum_{m=1}^{M}\sqrt{\eta_{mk}}\gamma_{mk})^{2}}{\splitfrac{1 + \rho_{d}\sum_{m=1}^{M}\eta_{mk}\gamma_{mk}\beta_{mk} + \rho_{d}\sum_{k'\neq k}\Big(\sum_{m=1}^{M}\eta_{mk'}\beta_{mk}||\mathbf{a}_{mk'}||_{2}^{2} + } {\tau\rho_{p}(|\sum_{m=1}^{M}\sqrt{\eta_{mk'}}\beta_{mk}\boldsymbol{\psi}_{k}^{H}\mathbf{a}_{mk'}|^{2} + \sum_{m=1}^{M}\eta_{mk'}\sum_{j=1}^{K}\beta_{mk}\beta_{mj}|\mathbf{\psi}_{j}^{H}\mathbf{a}_{mk'}|^{2})\Big)}}
		}.
		\end{aligned}
		\end{equation}
		\vspace{-1mm}
	\end{figure*}
	
	In \cite{Rao_Internet_2019} and \cite{Rao_cellfree_2018} the authors formulated a quasi-convex  max-min SINR power control optimization problem, which can be solved by a bisection search.  However, this approach has too high complexity if  the number of APs is large. Hence a simpler power control algorithm is required. It was observed in \cite{Nayebi2017Precoding} that if we magically knew optimal normalized transmit power $p_{m}$ as  of $m$-th AP, which is equal to $\sum_{k=1}^{K}\eta_{mk}\gamma_{mk}$, then the power optimization problem can be formulated as 
	\begin{equation}
	\allowdisplaybreaks
	\begin{aligned}
	\label{eq:max_min_dl_orth_convex}
	&\max_{\boldsymbol{\eta}}\min_{k}\text{SINR}_{k}^{\text{orth}}(\boldsymbol{\eta}) = \frac{\rho_{d}(\sum_{m=1}^{M}\sqrt{\eta_{mk}}\gamma_{mk})^{2}}{1 + \rho_{d}\sum_{m=1}^{M}p_{m}^{\text{opt}}\beta_{mk}}\\
	&\qquad\qquad\text{s.t.} \sum_{k'=1}^{K} \eta_{mk'}\gamma_{mk'} = p_{m}^{\text{opt}}\\
	& \qquad\qquad\quad\;\eta_{mk'} \geq 0, m = 1,..., M, k' = 1,..., K.
	\end{aligned}
	\end{equation}
	which is a convex problem \cite{Nayebi2017Precoding} and has significant smaller complexity.   
	
	\subsection{Power Control using Neural Network}
	It is observed  in \cite{Nayebi2017Precoding} that there is an exponential relationship between $\beta_{m}^{\text{max}}$ and $p_{m}^{\text{opt}}$ where $\beta_{m}^{\text{max}} = \max_{k=1,.., K}\beta_{mk}$. A exponential regression can then be implemented to predict $p_{m}^{\text{opt}}, \forall m$. We denote the outputs of the exponential regression as $p_{m}(\beta_{m}^{\text{max}}), \forall m$ and these outputs can be substituted into (\ref{eq:max_min_dl_orth_convex}) to solve the convex problem. 
	
	Based on these observations, low complexity power control can be implemented as follows. First, $p_{m}^{\text{opt}}, \forall m$ are approximated using exponential regression based on $\beta_{m}^{\text{max}}, \forall m$ as in \cite{Nayebi2017Precoding}. The outputs, $p_{m}(\beta_{m}^{\text{max}}), \forall m$ are then used in (\ref{eq:max_min_dl_orth_convex}).  Lastly, the obtained power control coefficients are regarded as the solution for IoT systems.
	
	However, the performance achieved using $p_{m}(\beta_{m}^{\text{max}})$ is not close enough to the optimal performance achieved by $p_{m}^{\text{opt}}$ and the generality of this method is limited. The relationship between $\beta_{m}^{\text{max}}$ and $p_{m}^{\text{opt}}$ obtained for one network often cannot be applied to other networks. Moreover,  for some networks no exponential relationship can be found between $p_{m}^{\text{opt}}$ and $\beta_{m}^{\text{max}}$. 
	
\begin{figure*}
		\begin{center}
			\includegraphics [width=0.7\textwidth]{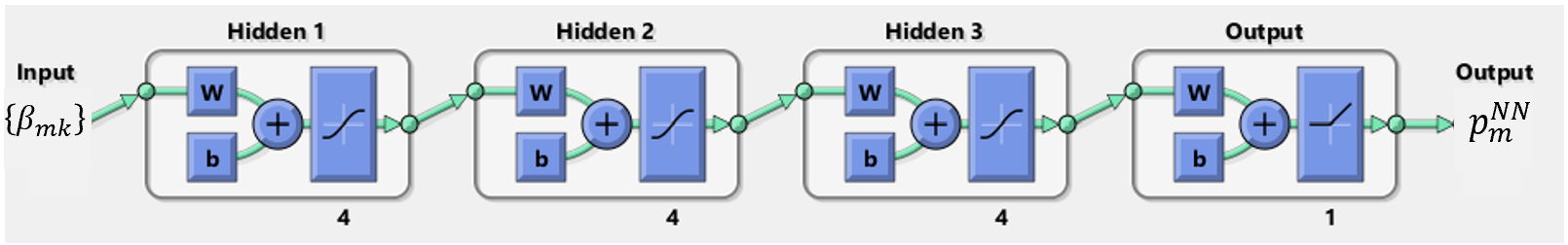}
			\caption{Neural network for predicting $p_{m}^{\text{NN}}, \forall m$.}\label{fig:NN}
		\end{center}
		\vspace{-6mm}
	\end{figure*}

	For this reason we propose to use NN to find $p_{m}^{\text{NN}}$ that would approximate $p_{m}^{\text{opt}}, \forall m$. A regression based NN shown in Fig. \ref{fig:NN} is used to generate  $p_{m}^{\text{NN}}, \forall m$ using known 
	$\{ \beta_{mk}\}$.  We used Levenberg-Marquardt (LM) training algorithm \cite{levenberg1944method}, \cite{marquardt1963algorithm} for this NN. With on the order of $10^4$ training samples, the training time is less than 1 hour in a typical laptop. It is also noted that the NN trained with LM algorithm has similar performance as the NN trained with higher complexity algorithms such as Bayesian Regularization.
	The experiments below are based on the same simulation settings as in Section \ref{Uplink Simulation Results}. 
	
\begin{figure}
		\begin{center}
			\centering \scalebox{0.9}{\includegraphics[width=\columnwidth]{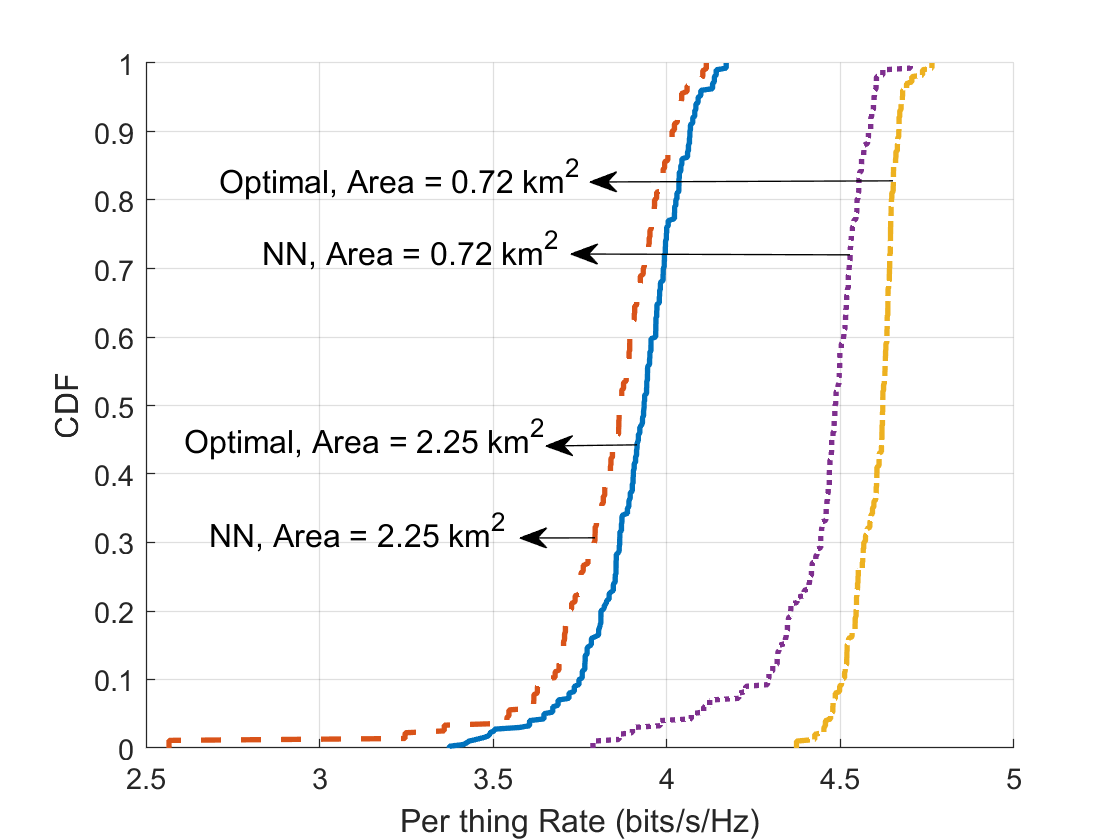}} \caption{$M = 128$, $K = 4$, $P_{d} = 200 \ \text{mW}$.}\label{fig:NN_Any_Area}
		\end{center}
		\vspace{-2mm}
	\end{figure}

	{\bf Experiment 1} We fix the numbers of APs and things, and train a NN for several areas. Next we reuse this NN for finding $p_{m}^{\text{NN}}$ for an area that was not used for training. In particular, we used the squares with areas $0.016, 0.063, 0.25, 0.56, 1, 4\ \text{km}^2$ for training our NN. Next we reuse it for the squares with areas $0.72$ and $2.25$ $\text{km}^2$, and plug in $p_m^\text{NN}$ into (\ref{eq:max_min_dl_orth_convex}). Results are shown in
	Fig.~\ref{fig:NN_Any_Area} where $P_d$ is the maximum transmit power of each AP. We see that the performance achieved with NN is quite close to the performance of the optimal power control. 
	
	\subsection{Scalable Power Control with High Energy Efficiency}
	For real life IoT networks, which involve a large number of APs, even the convex problem (\ref{eq:max_min_dl_orth_convex}) becomes too complex. Thus,  a scalable power control algorithm with very low complexity is required. In order to find such an algorithm we define the density of a network by
	\begin{equation}
	\label{eq:density_network}
	\text{Density} = \frac{\text{Number of APs}}{\text{Serving Area}}.
	\end{equation}
	
	Next we train NN for small areas with a given density and use obtained transmit powers $p_{m}^{\text{NN}}, \forall m$ for large areas with the same density.
	Further, instead of solving a convex optimization problem, we use uniform power control where the power control coefficients are given by
	\begin{equation}
	\label{eq:uniform_power_control}
	\eta_{m} = \frac{p_{m}^{\text{NN}}}{\sum_{k=1}^{K}\gamma_{mk}},\; \eta_{m} = \eta_{mk}, \forall k.
	\end{equation}
	It is important to note that $p_m^{\text{NN}},\ \forall m$ can be efficiently obtained and (\ref{eq:uniform_power_control}) has almost zero online complexity. Therefore this algorithm is scalable for large networks. 
	
\begin{figure}
		\begin{center}
			\centering \scalebox{0.85}{\includegraphics[width=\columnwidth]{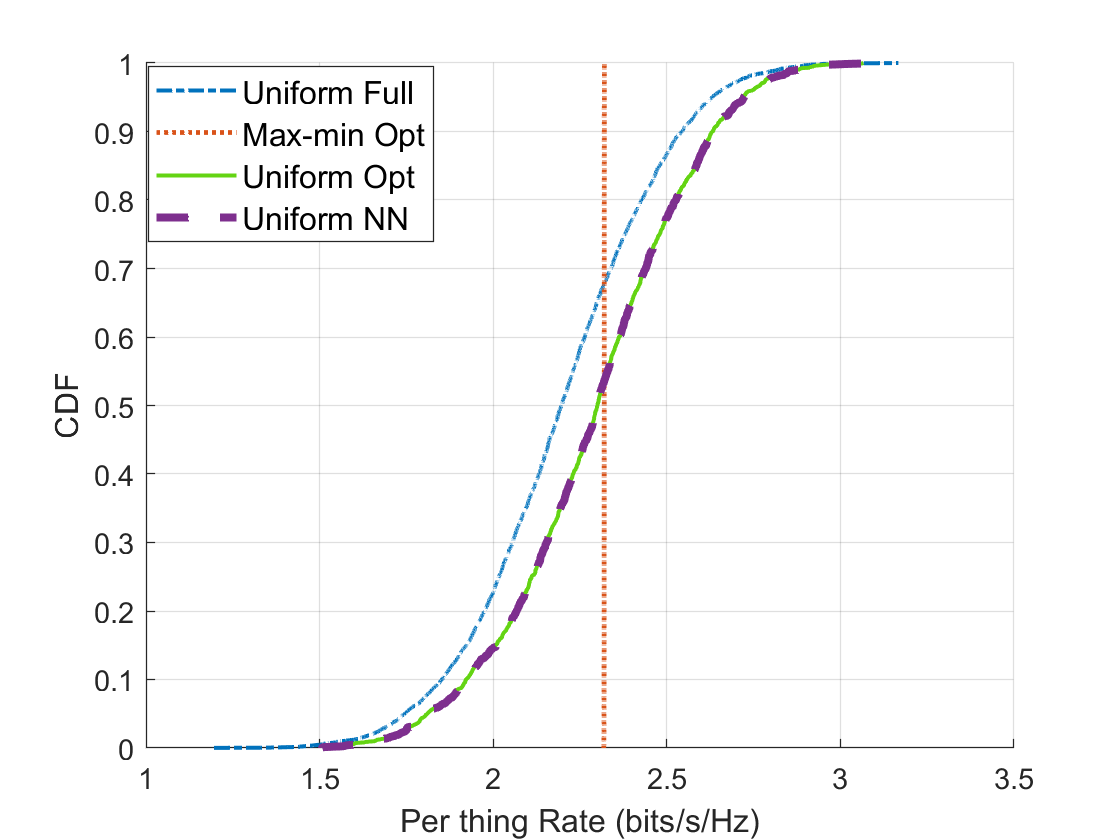}}\caption{ $M = 64$, $K = 16$, $P_{d} = 200 \ \text{mW}$, $\text{Area} = 0.03\ \text{km}^{2}$.}\label{fig:NN_fix_density}
		\end{center}
	\end{figure}

	{\bf Experiment 2} In this experiment, we fix  the area density and train NN with a small number of APs and things. We then use this NN to produce $p_{m}^{\text{NN}}$ for a large area with the same density, but possibly larger numbers of APs and things. 
	In Fig. \ref{fig:NN_fix_density} we use small $M$ and $K$ in order to compare our results with the true max-min optimal power control based on (\ref{eq:SINR_d_k_radom}) (Max-min Opt). The vertical appearance of “Max-min Opt” in Fig. \ref{fig:NN_fix_density} is because it effectively equalizes data rates for all users. Due to small area and wrap-around geometry, the difference between different realizations is very small. 
	We also present rates when we use (\ref{eq:uniform_power_control}) with maximal $p_m$ (Uniform Full), optimal $p_{m}^{\text{opt}}$  (Uniform Opt), and NN produced $p_{m}^{\text{NN}}$ (Uniform NN). We see that (\ref{eq:uniform_power_control}) with $p_{m}^{\text{NN}}$ and $p_{m}^{\text{opt}}$ produce basically identical results, though the complexity of finding $p_{m}^{\text{NN}}$ is much smaller.
	The gap between the rates obtained with maximal $p_m$ and $p_{m}^{\text{NN}}$ is not very large, however situation is different if we compare their energy efficiencies (see Fig. \ref{fig:NN_fix_density_large} below). 
	
	In Fig. \ref{fig:NN_fix_density_large} we use a large area with $M = 4096$ and $K = 1024$ (keeping the same area density). 
	We see that $p_{m}^{\text{NN}}$ provide very large energy gain, which is very important for IoT applications. Note that the energy efficiency defined for DL is $E_d = \sum_{k=1}^{K}R_{k}^{d}/\sum_{m=1}^{M}P_m$ where $P_m$ can be $P_d$ or $p_m^{\text{NN}}P_d$.
	\begin{figure}
		\begin{center}
			\centering \scalebox{0.85}{\includegraphics[width=\columnwidth]{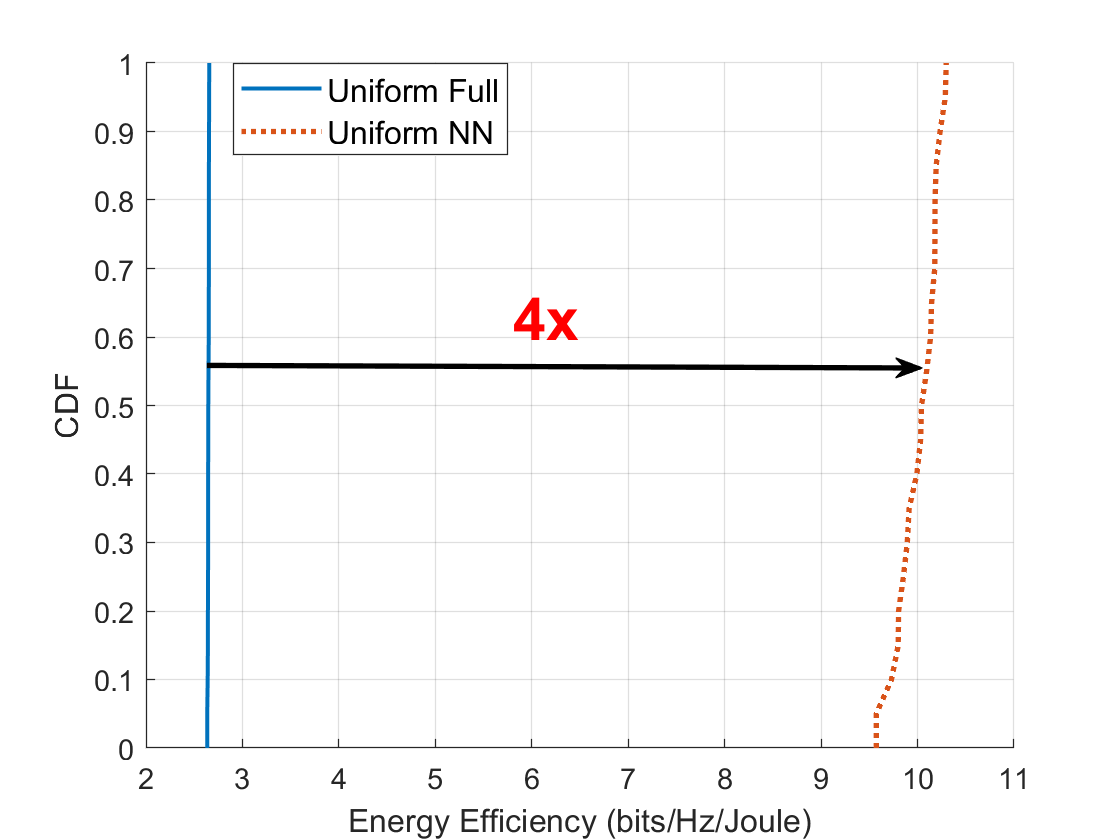}}
			\caption{$M = 4096$, $K = 1024$, $P_{d} = 200 \ \text{mW}$, $\text{Area} = 2\ \text{km}^{2}$.}\label{fig:NN_fix_density_large}
		\end{center}
		\vspace{-2mm}
	\end{figure}

	\section{Conclusion}
	In this work we proposed IoT supported by CF mMIMO with optimal components - LMMSE channel estimation and MMSE MIMO receiver. We derived a random matrix approximation for sensor's UL SINR and use it for efficient and low complexity power control algorithms that give large energy efficiency gains, which is very important for low powered sensors. 
	For DL transmission we proposed an NN based scalable algorithm for transmit power control. This algorithm, though sub-optimal, has almost zero online complexity and provides multifold gain in energy efficiency compared with full power transmission approach.  
	
	\bibliographystyle{IEEEtran}
	\bibliography{references}

\begin{thebibliography}{10}
\providecommand{\url}[1]{#1}
\csname url@samestyle\endcsname
\providecommand{\newblock}{\relax}
\providecommand{\bibinfo}[2]{#2}
\providecommand{\BIBentrySTDinterwordspacing}{\spaceskip=0pt\relax}
\providecommand{\BIBentryALTinterwordstretchfactor}{4}
\providecommand{\BIBentryALTinterwordspacing}{\spaceskip=\fontdimen2\font plus
\BIBentryALTinterwordstretchfactor\fontdimen3\font minus
  \fontdimen4\font\relax}
\providecommand{\BIBforeignlanguage}[2]{{%
\expandafter\ifx\csname l@#1\endcsname\relax
\typeout{** WARNING: IEEEtran.bst: No hyphenation pattern has been}%
\typeout{** loaded for the language `#1'. Using the pattern for}%
\typeout{** the default language instead.}%
\else
\language=\csname l@#1\endcsname
\fi
#2}}
\providecommand{\BIBdecl}{\relax}
\BIBdecl

\bibitem{Liu_Conver_Analysis}
L.~{Liu}, C.~{Yuen}, Y.~L. {Guan}, Y.~{Li}, and Y.~{Su}, ``{Convergence
  Analysis and Assurance for Gaussian Message Passing Iterative Detector in
  Massive MU-MIMO Systems},'' \emph{IEEE Transactions on Wireless
  Communications}, vol.~15, no.~9, pp. 6487--6501, 2016.

\bibitem{Liu_MassiveConnec_2018}
L.~{Liu} and W.~{Yu}, ``{Massive Connectivity With Massive MIMO—Part I:
  Device Activity Detection and Channel Estimation},'' \emph{IEEE Transactions
  on Signal Processing}, vol.~66, no.~11, pp. 2933--2946, 2018.

\bibitem{Zhang_UAD_CE}
Z.~{Zhang}, Y.~{Li}, C.~{Huang}, Q.~{Guo}, L.~{Liu}, C.~{Yuen}, and Y.~L.
  {Guan}, ``{User Activity Detection and Channel Estimation for Grant-Free
  Random Access in LEO Satellite-Enabled Internet-of-Things},'' \emph{IEEE
  Internet of Things Journal}, pp. 1--1, 2020.

\bibitem{Hoydis_13_UL/DL}
J.~Hoydis, S.~ten Brink, and M.~Debbah, ``Massive {MIMO} in the {UL/DL} of
  cellular networks: How many antennas do we need?'' \emph{IEEE Journal on
  Selected Areas in Communications}, vol.~31, no.~2, pp. 160--171, Feb. 2013.

\bibitem{Wagner_12_MISO}
S.~{Wagner}, R.~{Couillet}, M.~{Debbah}, and D.~T.~M. {Slock}, ``Large system
  analysis of linear precoding in correlated {MISO} broadcast channels under
  limited feedback,'' \emph{IEEE Transactions on Information Theory}, vol.~58,
  no.~7, pp. 4509--4537, July 2012.

\bibitem{silverstein1995empirical}
J.~W. {Silverstein} and Z.~D. {Bai}, ``On the empirical distribution of
  eigenvalues of a class of large dimensional random matrices,'' \emph{Journal
  of Multivariate Analysis}, vol.~54, no.~2, pp. 175--192, 1995.

\bibitem{Yan2020scalable}
\BIBentryALTinterwordspacing
H.~{Yan}, A.~{Ashikhmin}, and H.~{Yang}, ``{A Scalable and Energy Efficient IoT
  System}.'' [Online]. Available: \url{https://arxiv.org/abs/2005.06696}
\BIBentrySTDinterwordspacing

\bibitem{Ngo_17_Cellfree}
H.~Q. Ngo, A.~Ashikhmin, H.~Yang, E.~G. Larsson, and T.~L. Marzetta,
  ``Cell-free massive {MIMO} versus small cells,'' \emph{IEEE Transactions on
  Wireless Communications}, vol.~16, no.~3, pp. 1834--1850, Mar. 2017.

\bibitem{Tang_01_Pathloss}
A.~Tang, J.~Sun, and K.~Gong, ``Mobile propagation loss with a low base station
  antenna for {NLOS} street microcells in urban area,'' in \emph{IEEE 53rd
  Vehicular Technology Conference (VTC-Spring)}, 2001, pp. 333--336.

\bibitem{3GPP_ETSI}
3GPP, ``Digital cellular telecommunication system (phase 2+); radio network
  planning aspects,'' 3GPP, ETSI TR, 2010.

\bibitem{Wang08_Joint_Shadow}
Z.~Wang, E.~K. Tameh, and A.~R. Nix, ``Joint shadowing process in urban
  peer-to-peer radio channels,'' \emph{IEEE Transactions on Vehicular
  Technology}, vol.~57, no.~1, pp. 52--64, Jan. 2008.

\bibitem{Rao_Internet_2019}
S.~{Rao}, A.~{Ashikhmin}, and H.~{Yang}, ``{Internet of Things Based on
  Cell-Free Massive MIMO},'' in \emph{2019 53rd Asilomar Conference on Signals,
  Systems, and Computers}, 2019, pp. 1946--1950.

\bibitem{Rao_cellfree_2018}
------, ``{Cell-Free Massive MIMO with Nonorthogonal Pilots for Internet of
  Things},'' \emph{Bell Labs Report}, 2018.

\bibitem{Nayebi2017Precoding}
E.~{Nayebi}, A.~{Ashikhmin}, T.~L. {Marzetta}, H.~{Yang}, and B.~D. {Rao},
  ``{Precoding and Power Optimization in Cell-Free Massive MIMO Systems},''
  \emph{IEEE Transactions on Wireless Communications}, vol.~16, no.~7, pp.
  4445--4459, July 2017.

\bibitem{levenberg1944method}
K.~Levenberg, ``A method for the solution of certain non-linear problems in
  least squares,'' \emph{Quarterly of applied mathematics}, vol.~2, no.~2, pp.
  164--168, 1944.

\bibitem{marquardt1963algorithm}
D.~W. Marquardt, ``An algorithm for least-squares estimation of nonlinear
  parameters,'' \emph{Journal of the society for Industrial and Applied
  Mathematics}, vol.~11, no.~2, pp. 431--441, 1963.

\end{thebibliography}
\end{document}